\documentclass{osa-article}
\usepackage{bm}
\usepackage{fancyhdr}
\newcommand{\argmin}{\mathop{\rm argmin}\limits}

\journal{oe}
\articletype{Research Article}
\begin{document}

\title{Wiener-Hammerstein model and its learning for nonlinear digital pre-distortion of optical transmitters}

\author{Takeo Sasai,\authormark{1,*} Masanori Nakamura,\authormark{1} Etsushi Yamazaki, \authormark{1} Asuka Matsushita, \authormark{1} Seiji Okamoto, \authormark{1}  Kengo Horikoshi, \authormark{1}  and Yoshiaki Kisaka\authormark{1}}

\address{\authormark{1}NTT Network Innovation Labs, NTT Corporation, 1-1 Hikari-no-oka, Yokosuka, Kanagawa, 239-0847 Japan}
% \authormark{2}NTT Electronics Corporation, 1-1-32 Shin-Urasima-cho, Yokohama, Kanagawa-ku, Kanagawa, 221-0031 Japan}

\email{\authormark{*}takeo.sasai.cp@hco.ntt.co.jp} %% email address is required



\begin{abstract}
We present a simple nonlinear digital pre-distortion (DPD) of optical transmitter components, which consists of concatenated blocks of a finite impulse response (FIR) filter, a memoryless nonlinear function and another FIR filter. The model is a Wiener-Hammerstein (WH) model and has essentially the same structure as neural networks or multilayer perceptrons. This awareness enables one to achieve complexity-efficient DPD owing to the model-aware structure and exploit the well-developed optimization scheme in the machine learning field. The effectiveness of the method is assessed by electrical and optical back-to-back (B2B) experiments, and the results show that the WH DPD offers a 0.52-dB gain in signal to noise ratio (SNR) and 6.0-dB gain in optical modulator output power at a fixed SNR over linear-only DPD.
\end{abstract}

%%%%%%%%%%%%%%%%%%%%%%%%%%  body  %%%%%%%%%%%%%%%%%%%%%%%%%%
\section{Introduction}
The growing demand in throughput in data centers has prompted studies on high spectral efficiency (SE) optical transmission \cite{chen201916384,olsson2018record,sasai2020laser}. Such systems use high order quadrature amplitude modulation (QAM), which imposes severe requirements on the received SNR. In short-reach systems, transceiver impairments such as limited bandwidth and in-out nonlinearity of the components are the main performance-limiting factors of the received SNR. Especially in unamplified systems that the 400ZR project in the Optical Internetworking Forum (OIF) addresses for low-cost coherent interfaces \cite{OIF}, insufficient optical power at receivers (Rxs) is one of the key issues \cite{zhang2012design}, and the high modulator output power at transmitters (Txs) is a strong requirement. However, increasing Tx ouput power is accompanied by an SNR penalty due to the nonlinear saturation of modulator driver amplifiers (DA) and sinusoidal nonlinearity in Mach-Zehnder modulators (MZMs).

Various DPD approaches have been introduced to pre-compensate for the frequency response and the nonlinearities of the analog components. One common approach is the Volterra series or its truncated versions, such as the memory polynomial, which can be applied to a wide variety of nonlinear systems with memory \cite{morgan2006generalized, berenguer2015nonlinear, faig2019dimensions, sadot2019digital}. The look-up-tables (LUT) are also major tools for the correction of sample or symbol nonlinear deviation \cite{zhalehpour2019experimental, ke2014400}. Machine learning methods have also attracted much attention for their potential to extract nonlinear properties of the devices well, and neural networks \cite{paryanti2020direct} and the extreme learning machines \cite{schaedler2019ai} have been applied to nonlinear DPD. These are powerful tools for nonlinearity mitigation but tend to be black-boxes, are not fully complexity-efficient, and make it difficult to interpret the learned filters. Other approaches are based on more structure-aware models, such as the Wiener and the Hammerstein model, which have been intensively studied in the field of system identification \cite{schoukens2017identification,gilabert2005wiener}. The Wiener-Hammerstein (WH) (or Hammerstein-Wiener, HW) model, which has the concatenated structure of a linear time invariant system, and static nonlinearity, and another linear system, is suitable for systems with multiple series-connected components \cite{schoukens2017identification}.

Interestingly, these concatenated models have a similar or essentially the same structure as neural networks or multilayer perceptrons. This similarity enables one to benefit from both the behavioral model in system identification and powerful tools in machine learning as following: (i) the behavioral model allows efficient compensation compared to completely black-box methods such as fully-connected or other types of neural networks; (ii) physical interpretation and insights into the learning and the devices of interest are available from meaningful parameters of the learned model; (iii) the machine learning aspect enables better performance by leveraging various optimization techniques such as stochastic gradient descent based on backpropagation \cite{hecht1992theory} and updating schemes (e.g. Adagrad \cite{duchi2011adaptive} and Adam \cite{kingma2014adam}); (iv) simultaneous optimization of nonlinear and linear coefficients enables discriminative measurement of the linear responses before and after the nonlinearity (and vice versa) \textit{at once} due to the non-commutativity of nonlinear and linear phenomena. Similar ideas can be seen in the context of fiber nonlinearity compensation \cite{hager2018nonlinear} and the estimation of longitudinal fiber loss and dispersion profiles \cite{sasai2020simultaneous}.

\begin{figure}[b]
\centering\includegraphics[width=10cm]{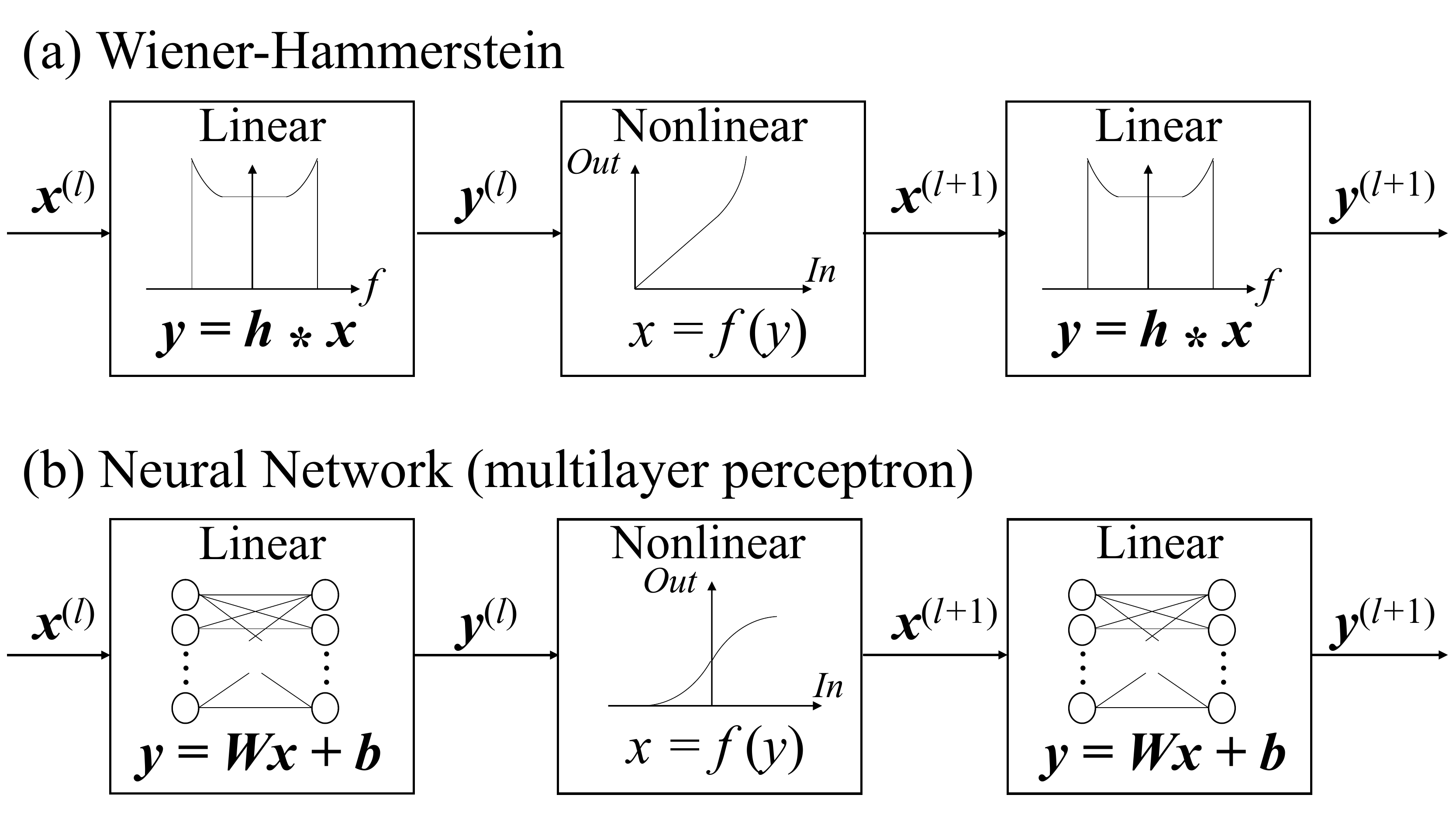}
\caption{Block diagrams of (a) Wiener-Hammerstein model and (b) neural network.}
\label{WHandNN}
\end{figure}

In this paper, we introduce a simple nonlinear DPD model for DA nonlinearity and the overall Tx frequency response based on the WH model, which consists of cascaded blocks of an instantaneous nonlinear function ($f(x) = x + ax^{3}$) sandwiched between two FIR filters. We provide the optimization scheme of the model, inspired by the similarity between neural-network structure and the WH model as shown in Fig. \ref{WHandNN}. We conducted electrical and optical B2B experiments to assess the effectiveness of the model and demonstrate a 0.52-dB gain at the maximum received SNR and 6.0-dB gain in modulator output power over linear pre-distortion only. We also discuss that simply applying pre-distortion is insufficient and optimizing the operating point (i.e., input amplitude adjustment) is required due to the increased input peak-to-average-power ratio (PAPR) (or back-off). This is a general problem and not limited to WH pre-distorter but also applicable to other nonlinear DPD approaches.

The paper is organized as follows. In Section 2, we describe the WH model for the Tx DPD and the coefficient learning scheme. In Sections 3 and 4, we discuss the assessment of the effectiveness of the model conducted through electrical and optical B2B experiments, respectively.

\section{Wiener-Hammerstein model and its learning}
\subsection{Wiener-Hammerstein model for optical transmitters}
The transmitter of optical transmission systems is typically composed of a digital-to-analog converter (DAC), DAs, and a MZM, and each component has frequency responses and in-out nonlinear characteristics. Combining these components forms the cascaded multilayer of nonlinearities and linear time invariant systems. This structure can be interpreted as a cascaded or generalized WH (or HW) model.

On the basis of this model, the complete DPD has to be in reverse order, which is also a generalized WH (or HW), and each inverse characteristic block cancels out the corresponding actual device impairment. In this paper, however, we focus on pre-distortion of DA nonlinearity and frequency responses of components (DAC, DA, and MZM). This model is a simple WH model (Fig. \ref{WHandNN} (a)), consisting of a memoryless nonlinear function sandwiched between two FIR filters (linear, nonlinear, and linear, LNL).Note that we can extend the WH DPD model to a complete one that includes the DAC and MZM nonlinearity by simply adding the layers of the WH structure (LNLNL...). As shown in Fig.  \ref{WHandNN}, the structure of the WH model is quite similar to or essentially the same as that of neural networks (Fig. \ref{WHandNN} (b)), which also have the concatenated blocks of (linear) Affine transformation and (nonlinear) activation functions. Hence, we can exploit benefits introduced in Section 1 from both structure-aware modeling and machine learning.

One might argue that one FIR filter is sufficient for the compensation of the frequency response in transmitters and the Wiener model (one linear and one nonlinear filters) is enough to compensate for the transmitter nonlinearity. However, DAs generally have the frequency-dependent nonlinearity as analyzed in \cite{singerl2007constructing}. In terms of the input voltage, the start of saturation varies with frequencies. The same thing can be said in terms of the output voltage since its maximum (saturated) amplitudes are also different among different frequencies. This implies that frequency responses exist in drivers both before and after the saturation nonlinearity. Furthermore, actual transmitters have DAC, DAs, and MZM, and the DA nonlinear response is sandwiched by linear responses. Since linear and nonlinear phenomena are non-commutative, the compensation order must be the reverse of the actual systems, which is the WH model (LNL).

\subsection{Learning algorithm}
\begin{figure}[htbp]
\centering\includegraphics[width=12cm]{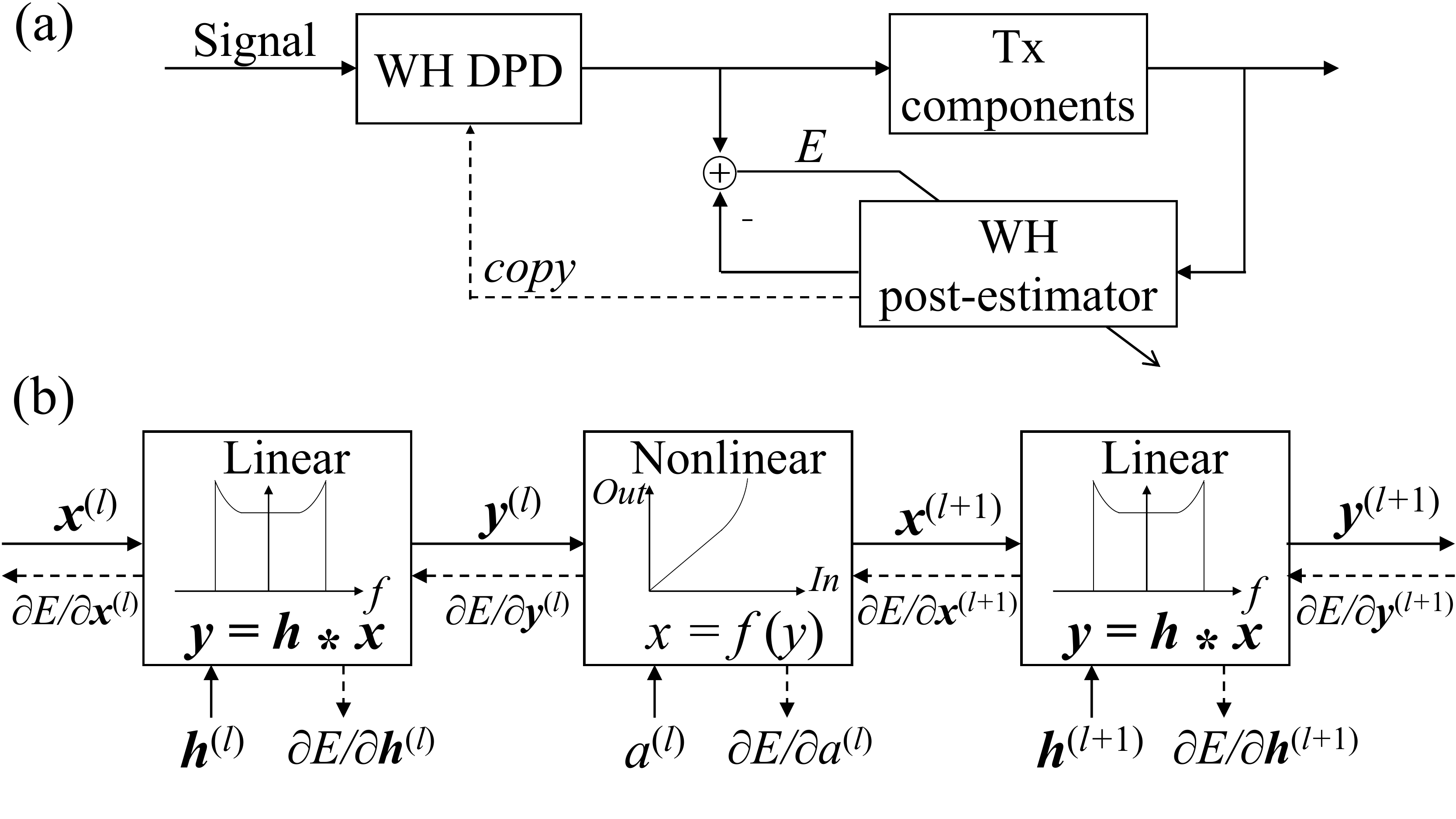}
\caption{(a) Indirect learning architecture for the WH DPD. (b) Calculation flow of partial derivatives of $E$ with respect to WH model coefficients.}
\label{IDLandBP}
\end{figure}

In \cite{pan2011wiener}, the WH post-estimator has been used in the context of the fiber nonlinearity compensation and the algorithm has been based on the one proposed in \cite{hegde2002series}, which is derived for LNL systems. The optimization scheme described here assumes the generalized WH DPD for generality, which can be used for LNLNLN... systems.
The overall learning architecture is based on indirect learning \cite{eun1997new} as depicted in Fig. \ref{IDLandBP} (a).

The learning is the fully-supervised learning using the transmitted waveform. Non-DPD signals are generated and passed through the Tx components. The received signals are then fed into the post-estimator, which later becomes the pre-distorter. In the estimation stage, linear parts, which work as the frequency response compensation, can be written as
\begin{equation}
\label{linearConv}
\bm{y}^{(l)} = \bm{h}^{(l)} \ast \bm{x}^{(l)},
\end{equation}
where
\begin{math}
\bm{x}^{(l)} = (x_{0}^{(l)},...x_{n}^{(l)},...x_{N}^{(l)})^{T}
\end{math}
is the $N$ input samples of the $l$-th layer, 
\begin{math}
\bm{y}^{(l)}
\end{math}
the output samples, 
\begin{math}
\bm{h}^{(l)} = (h_{0}^{(l)},...h_{k}^{(l)},...h_{K}^{(l)})^{T}
\end{math}
the $K$-tap filter coefficients, and $\ast$ denotes convolution. The $l$-th layer ($l=1,2,...,L$) means the $l$-th set of linear and nonlinear functions, and $L$ denotes the total number of linear and nonlinear sets. In our experiments, $L$ = 2 is used and $L$-th (second) layer contains only a linear filter to form a simple WH model. Subsequently, the outputs $\bm{y}^{(l)}$go through memoryless nonlinear function
\begin{equation}
\label{NLfunc}
x_{n}^{(l+1)} = f(y_{n}^{(l)}),
\end{equation}
where $f$ is bijective and differentiable. $f^{-1} = \arctan \quad (f^{-1}: \mathbb{R} \rightarrow [-\pi/2, \pi/2])$ \; is the most widely used model for DA saturation, but in this paper we use 
\begin{equation}
\label{NLfuncSpec}
f(y_{n}^{(l)}) = y_{n}^{(l)} + a \left( y_{n}^{(l)} \right) ^{3},
\end{equation}
because (i) we can avoid the input $y_{n}^{(l)}$ going out of the domain $[-\pi/2, \pi/2]$ in the iterative learning process, (ii) the reduced complexity is preferable, and (iii) the third order coefficient of Taylor series expansion of $f = \tan$ is the most significant among nonlinear basis functions. Then, the iterative process of linear and nonlinear operations continues until the last $L$-th layer. The task is to find the optimum values for all linear and nonlinear coefficients $\hat{h}_{k}^{(l)}, \hat{a}^{(l)}$ that minimize cost function $E$ as
\begin{equation}
\label{argmin}
\hat{h}_{k}^{(l)}, \hat{a}^{(l)} = \argmin E
= 
\argmin \left[\frac{1}{2}\|\bm{y}^{(L)} - \hat{\bm{y}}\|^{2}\right],
\end{equation}
where $\hat{\bm{y}}$ is the reference samples (i.e., the transmitted signal). One can add a regularization term in the cost function for better convergence.  This is a least square problem for parameters of cascaded blocks and can be solved by the backpropagation algorithm, one of the gradient descent methods as depicted in Fig. \ref{IDLandBP} (b). The partial derivatives $\partial E / \partial h_{k}^{(l)}$ and $\partial E /\partial a^{(l)}$
are calculated for coefficient updates using the total derivative and the chain rule. The derivatives of $E$ with respect to the input of $E$, $\bm{y}^{(L)}$ are
\begin{equation}
\label{firstDerivative}
\frac{\partial E}{\partial \bm{y}^{(L)}} = \bm{y}^{(L)} - \hat{\bm{y}}.
\end{equation}
In linear blocks, the partial derivatives are obtained as
\begin{equation}
\label{LinearDerivativeH}
\frac{\partial E}{\partial \bm{h}^{(l)}} = \frac{\partial E}{\partial \bm{y}^{(l)}} \ast flip(\bm{x}^{(l)}),
\end{equation}
\begin{equation}
\label{LinearDerivativeX}
\frac{\partial E}{\partial \bm{x}^{(l)}} = \frac{\partial E}{\partial \bm{y}^{(l)}} \ast flip(\bm{h}^{(l)}),
\end{equation}
where $flip(\cdot)$ is the reversing operation in the order of the elements. Eq. (\ref{LinearDerivativeX}) means that the backpropagation of the convolution is expressed as the convolution of the back-propagated derivative and the flipped filter. Then, $\partial E / \partial \bm{x}^{(l)}$ is back-propagated to the previous nonlinear blocks. In nonlinear blocks, the same operation is conducted as
\begin{equation}
\label{NLDerivativeA}
\frac{\partial E}{\partial a^{(l)}} = \sum_{n=1}^{N} \frac{\partial E}{\partial x_{n}^{(l+1)}} \frac{\partial x_{n}^{(l+1)}}{\partial a^{(l)}} =  \sum_{n=1}^{N} \frac{\partial E}{\partial x_{n}^{(l+1)}} \left( y_{n}^{(l)} \right)^{3},
\end{equation}
\begin{equation}
\label{NLDerivativeY}
\frac{\partial E}{\partial y_{n}^{(l)}} = \frac{\partial E}{\partial x_{n}^{(l+1)}} \frac{\partial x_{n}^{(l+1)}}{\partial y_{n}^{(l)}} = \frac{\partial E}{\partial x_{n}^{(l+1)}} \left[ 1 + 3a^{(l)} \left( y_{n}^{(l)} \right) ^{2} \right],
\end{equation}
and again  $\partial E / \partial \bm{y}^{(l)}$ is back-propagated to another linear block. Finally, we update the coefficients from the obtained partial derivatives $\partial E / \partial h_{k}^{(l)}$ and $\partial E /\partial a^{(l)}$ . In this paper, we use Adam \cite{kingma2014adam} as the updating algorithm for better convergence of the coefficients.

One of the challenges in the estimation of WH model is to estimate different linear filters separately \cite{schoukens2014fast}. (This is also the case with different nonlinear functions in a generalized WH model.) To address this problem, the proposed approach exploits two facts: (i) all linear and nonlinear phenomena in the transmitters are non-commutative, which means the signal cannot be completely compensated with the \textit{wrong} order of the linear filters and nonlinear functions and there must be a correct order that gives the global optimum; and (ii) the optimization scheme presented herein is a simultaneous optimization of all linear and nonlinear coefficients $h_{k}^{(l)}, a^{(l)}$. If the estimation is not simultaneous such as small-signal measurement and low-frequency measurement \cite{berenguer2015nonlinear}, the estimated responses will be the combined version of each response and not be separated. As a different example of distinguishing cascaded linear response, the estimation of fiber nonlinearity and dispersion (linear response) profile along the transmission fiber has successfully demonstrated in \cite{sasai2020simultaneous} on the basis of the simultaneous learning of the nonlinear and linear coefficients in nonlinear Schr\"{
o}dinger equation, which is also composed of concatenated nonlinear and linear operations.

\subsection{Computational complexity}
The computational complexity of the nonlinear DPD is a crucial parameter especially when implementing in an application specific integrated circuit (ASIC). In the generalized WH model, the complexity can be derived easily because only FIR filters and memoryless nonlinear functions are adopted. For FIR filters with $K$ taps, the multiplications per sample are $K$, and the additions per sample are $K-1$. For nonlinear functions, let us assume $M$-th polynomial functions $f(x) = x + \sum_{m=2}^{M}a_{m}x^{m}$ for generality. The number of multiplications for $m$-th order term is $m$ and thus $\sum_{m=2}^{M}m = \frac{1}{2}M(M+1) - 1$ for total, while the additions are $M-1$. Then, we can add up these for all linear and nonlinear blocks in the generalized WH model. In our experiment, we only use a simple WH model (two FIR filters with $K_{1}$,$K_{2}$ taps and the third-order memoryless nonlinear functions ($M=3$) without the second order term). Thus, the total multiplications and additions per sample are $K_{1} + K_{2} + 3$ and $(K_{1}-1) + (K_{2}-1) + 1 = K_{1} + K_{2} - 1$, respectively.

\section{Electrical B2B experiment}
\subsection{Experimental setup}
We first assessed the effectiveness of the WH DPD through electrical B2B experiments and obtained the coefficients of the model for DAC and DA. In the Tx offline digital signal processing (DSP), a pseudo random bit sequence is generated using the Mersenne twister and mapped into 16QAM symbols with a length of 65,536. Nyquist-pulse shaping is applied with a root-raised-cosine filter with a roll-off factor of 0.2. The signal is set to 32 GBd and operated at 2 Sa/symbol. DPD is carried out once the coefficients are estimated. The signals are resampled to 120-GSa/s to be emitted from a 40-GHz arbitrary waveform generator (AWG) with a 8-bit resolution, which works as a DAC. No intentional clipping and other quantizations are performed. The output amplitude varies from 0.10 to 0.80 Vpp and is amplified by 60-GHz DAs, whose gain is fixed at 22 dB. A 110-GHz ADC with a 10-bit resolution digitizes the incoming signals at 256 GSa/s, which are then resampled to 64 GSa/s. After the signals are synchronized to the reference data, the learning of the WH model begins. The indirect learning is performed only once. The two FIR filters have 401 taps to sufficiently compensate for the reflection of electrical wiring between the DAC and DA, and the DA and ADC. Their initial values are set to the unit impulse filter for two FIR filters and $a$ = 0 for the nonlinear function. After the coefficients are converged, the estimated values and maximum input amplitude of the nonlinear block in the WH model are stored and moved to the DPD part in the transmitter. The signal amplitude is scaled to the stored value before input to the nonlinear function to simulate the same nonlinearity as that in the Rx estimator.

\begin{figure}[t]
\centering\includegraphics[width=12cm]{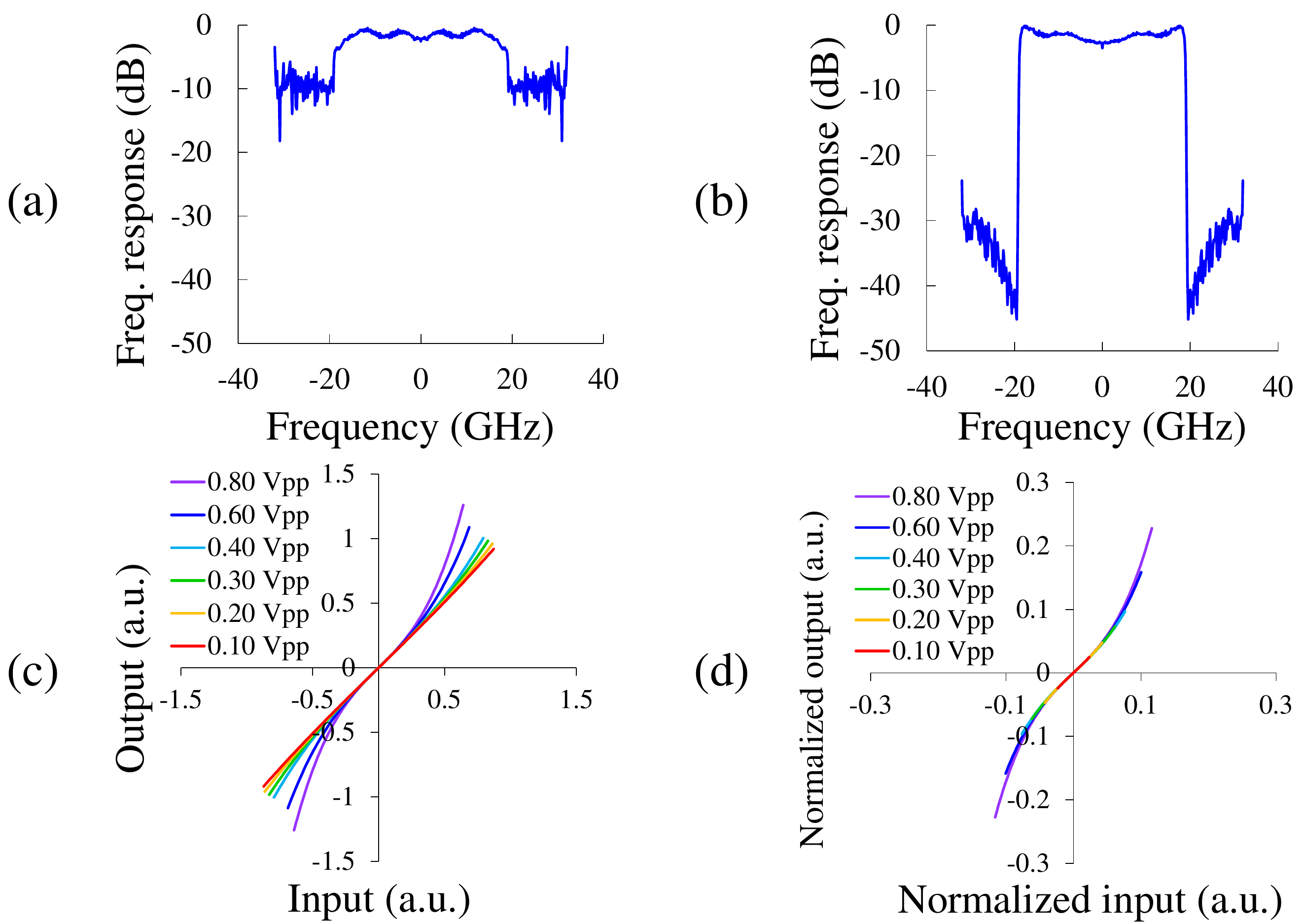}
\caption{Estimated DA nonlinearity and frequency responses before and after saturation based on the WH model. (a) First and (b) second FIR filter. Instantaneous nonlinear function ($f(x) = x + ax^{3}$) (c) without and (d) with amplitude normalization. }
\label{estimatedResponse}
\end{figure}

\subsection{Estimated WH model}
The estimated responses of the two linear filters and the nonlinear function are presented in Fig. \ref{estimatedResponse}. The first FIR filter in Fig. \ref{estimatedResponse} (a) corresponds to the combined frequency response of actual devices after DA nonlinearity (DA package, electrodes, and ADC) and the second one in Fig. \ref{estimatedResponse} (b) includes the DAC, electrodes and the DA. 

As for the in-band shape, the first FIR filter does not show peaks on the edge frequency while the second one does. This can be explained by organizing three facts: (i) a driver has different frequency responses between before and after the saturation nonlinearity as we have discussed in Section 2.1; (ii) the total frequency response of the driver is the overall (convoluted) frequency response of responses before and after saturation; (iii) the total frequency response of the dirver (bandwidth: 60 GHz) is almost flat in the signal bandwidth. From these facts, if the second FIR filter (Fig. 3(b)) shows the peaking on the edge frequency, the first FIR filter should show the reverse. That is why we observe the dropping on the edge frequency in Fig. 3(a). 

The first FIR filter does not suppress the out-of-band noise as the second one does because the following nonlinear function requires the out-of-band harmonic components generated in the DA to fully mitigate nonlinearity. The estimated nonlinear function $f(x) = x + ax^{3}$ is shown in Fig. \ref{estimatedResponse} (c) as a scatter plot of the input and output of $f$. The estimation was conducted with different input amplitudes $V_{in}$ (from 0.10 to 0.80 Vpp) of the DA, and as $V_{in}$ increases, the third nonlinear coefficient $a$ becomes larger. Note that the input amplitudes of $f$ are similar values among different $V_{in}$ in Fig. \ref{estimatedResponse} (c) because the root mean square (RMS) of the signals is adjusted to that of the reference waveforms in advance of being filtered by the WH estimator. To obtain deeper insights, we multiplied the input and output amplitudes of $f$ in Fig. \ref{estimatedResponse} (c) by the received amplitudes in the ADC and scatter-plotted them in Fig. \ref{estimatedResponse} (d). All the curves fit one nonlinear curve well, which implies the memoryless nonlinear function is a possible candidate for the actual model. Furthermore, this means that only a one-time measurement of nonlinear coefficient $a$ is required. We do not need to re-estimate $a$ and can re-use it even when the input level of the DA varies.

\begin{figure}[t]
\centering\includegraphics[width=13cm]{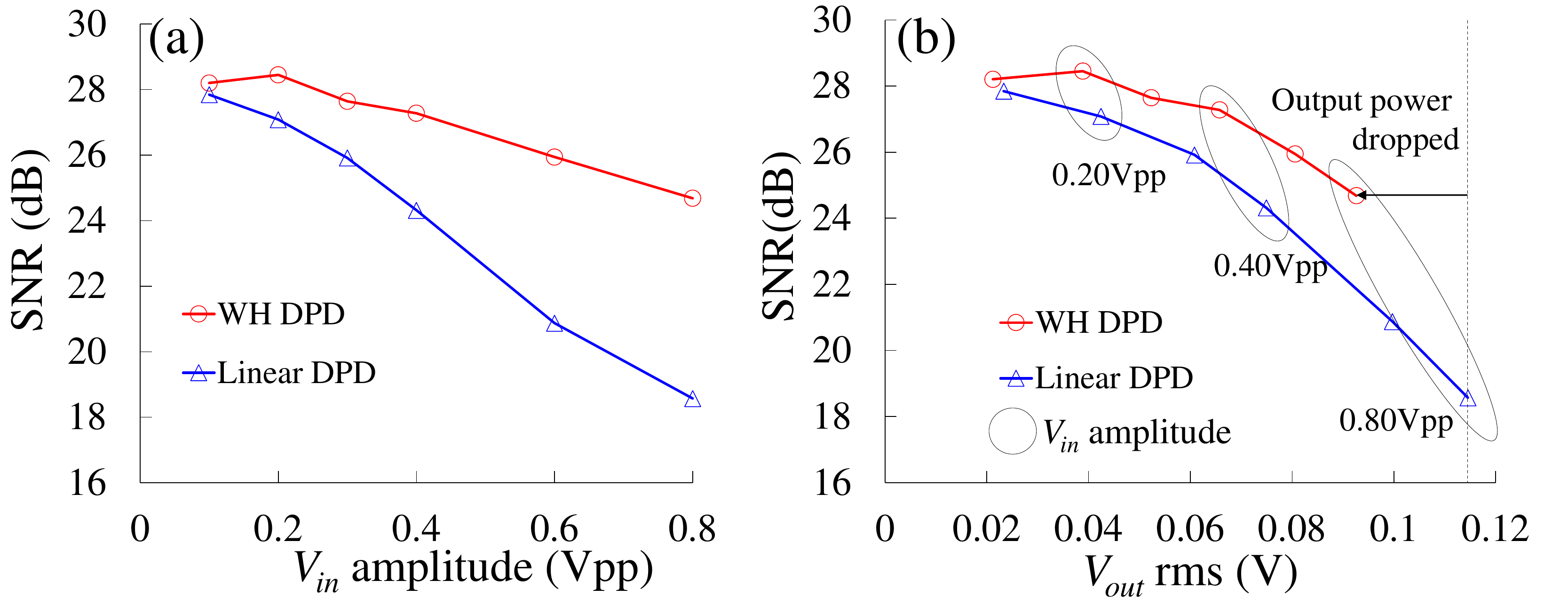}
\caption{Electrical B2B performance of the WH DPD and linear-only DPD. (a) SNR vs. DA input amplitude. (b) SNR vs. DA output RMS.}
\label{ResultElecB2B}
\end{figure}

\begin{figure}[t]
\centering\includegraphics[width=10cm]{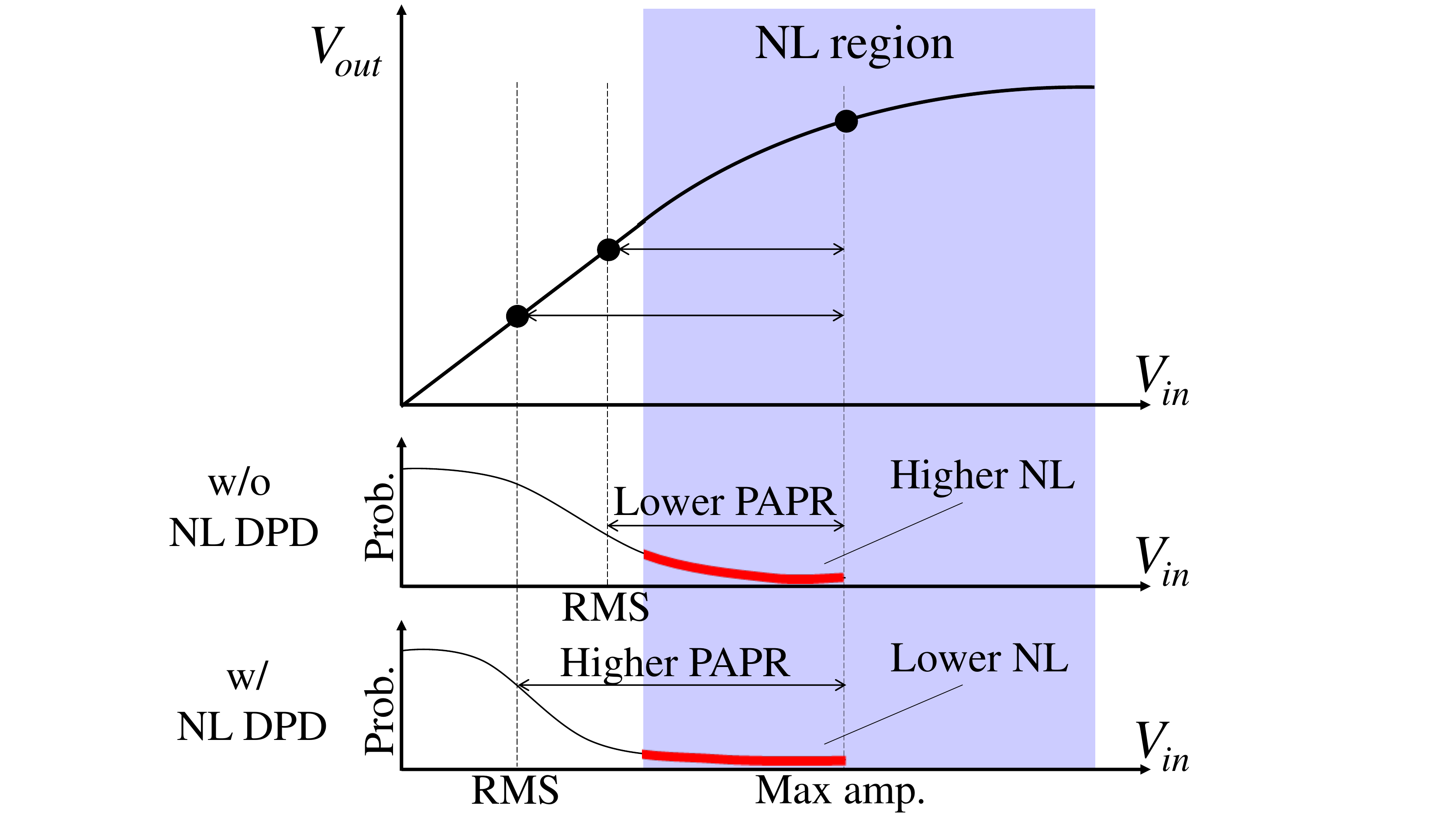}
\caption{Operation points of DA without and with NL DPD under a fixed amplitude.}
\label{backOff}
\end{figure}

\subsection{Performance of WH DPD in electrical B2B}\label{Performance of WH DPD in electrical B2B}
Figure \ref{ResultElecB2B} shows the resulting performance of the WH DPD (red circle) and linear-only DPD (blue triangle). The SNR is calculated in two sample per symbol domain using SNR = 
$E\left[\|\bm{r}\|^{2}\right]/E\left[\|\bm{r} - \bm{d}\|^{2}\right]$, where $E$ denotes the mean operation, and $\bm{r}$ and $\bm{d}$ denotes the reference signals and the demodulated signals, respectively. The analysis is conducted in both input amplitude and output RMS perspectives as shown in Fig. \ref{ResultElecB2B} (a) and (b), respectively. When compared to the linear-only DPD in terms of input amplitude, the WH DPD seems to have an overwhelmingly higher performance. However, this is not fair because, as illustrated in Fig. \ref{backOff}, when the nonlinear DPD (NL DPD) is applied to the signal, fewer samples pass through the nonlinear saturation region of the DA than the linear-only DPD due to an increased PAPR (in other words, a larger back-off from the saturation point) under a fixed maximum input amplitude. Thus, NL DPD shows an unfairly higher performance than the linear-only DPD. In this case, the comparison should be performed from the output power perspective as in Fig. \ref{ResultElecB2B} (b). Only the horizontal axis is replaced with output ($V_{out}$ RMS) of DA. The followings are notable: (i) $V_{out}$ RMS reduction is observed when the WH DPD is applied in spite of the same input amplitude, which indicates that the DA actually operates with a larger back-off; and (ii) even when comparing output $V_{out}$ RMS, the WH DPD offers a higher received SNR, which clearly means effective nonlinearity mitigation. Problem (i) is applicable to other targeted components and NL DPD methods that enhance the PAPR of the signal.

The increase in PAPR due to NL DPD can also impose larger quantization noise at the DAC. In general, most NL DPD methods increases PAPR and suffer from the quantization noise. In our cases, we employed 16QAM and the results show the compensation gain, but the performance can possibly degrade when we choose higher order QAM signals, which are more vulnerable to the quantization. This is a general problem of NL DPD and the method presented in \cite{yoffe2019low} can be used to reduce the effect of the quantization noise.

\section{Optical B2B experiment}
\subsection{Experimental setup}
\begin{figure}[t]
\centering\includegraphics[width=12cm]{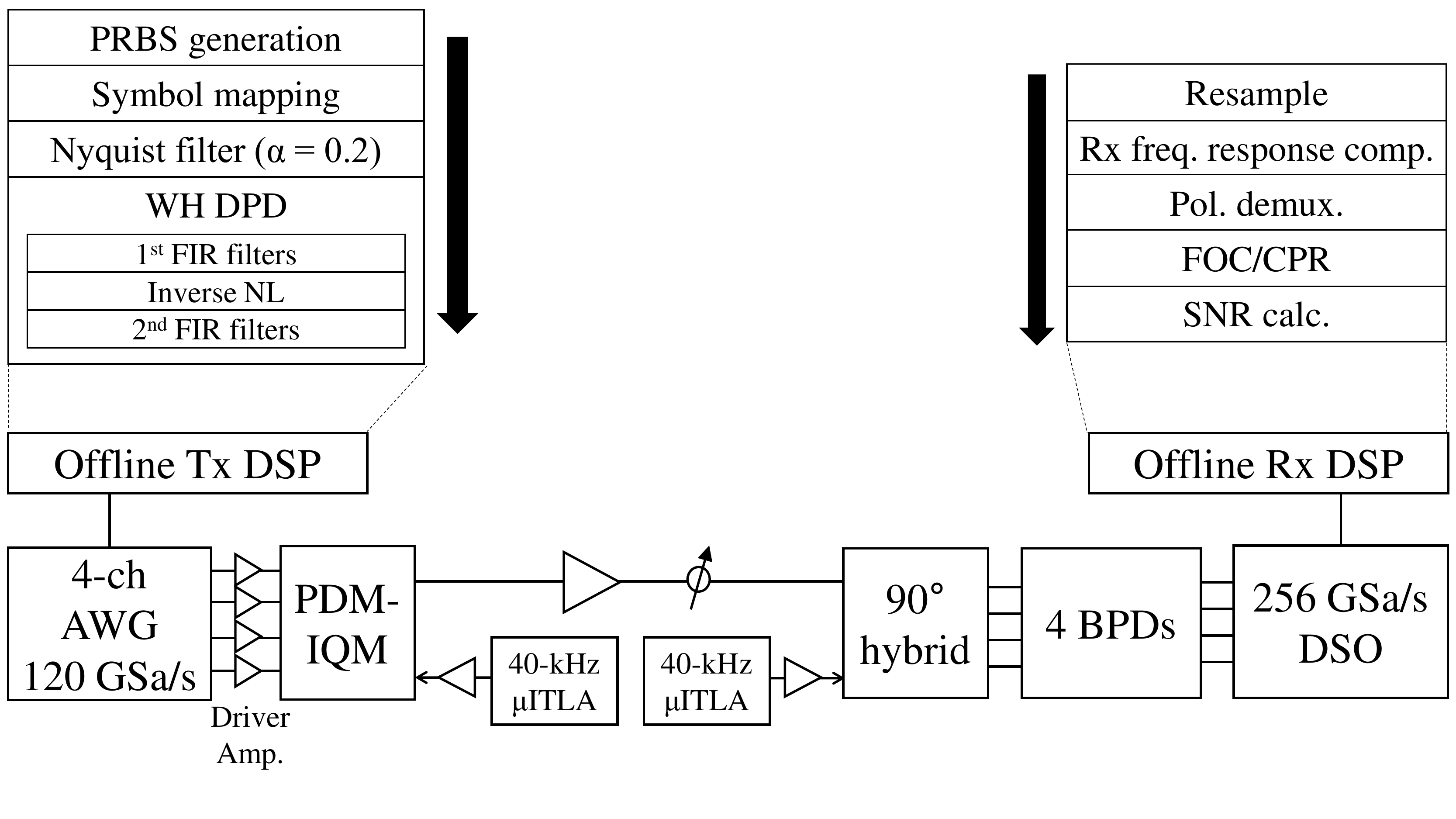}
\caption{Experimental setup for optical B2B measurement of WH DPD performance.}
\label{expSetup}
\end{figure}

We also evaluated the effectiveness of the WH DPD through optical B2B experiments. Figure \ref{expSetup} shows the experimental setup. The electrical part of the setup is the same as the configuration explained in Section 3. After emitted from the DAs, the signals are input to a 35-GHz dual polarization IQ-MZM. The laser wavelength is set to 1555.735 nm, and the typical linewidth is 40 kHz. The laser's output power is 9.3 dBm and amplified by a polarization-maintaining erbium doped fiber amplifier (EDFA). The input power and the insertion loss of the MZM are 20 dBm and 14 dB, respectively. The output power of the MZM is monitored and used for the horizontal axis of Fig. \ref{ResultOptB2B}. We always adjusted the output power of the IQ-MZM so that the powers of x and y polarization become the same. First, we turn on only the XI and XQ output of the MZM and measure the x-polarization power. The same operation is done for the y-polarization and the output amplitudes of the DAC are adjusted to be the same. An EDFA boosts the signals, which are then detected using a 90 degree hybrid and 100-GHz balanced photo diodes (BPDs). In Rx DSP, the Rx-side frequency response including the static amplitude and phase response is compensated first, and the signals go through a 61-tap T/2-spaced adaptive equalizer \cite{sano20155} with a butterfly configuration, which works as the polarization de-multiplexing and the residual and dynamic skew compensation. Then, the frequency offset compensation (FOC) and the carrier phase recovery (CPR) are performed using a second-order digital phase locked loop \cite{sasai2020laser}. Finally, we calculate SNR in one sample per symbol using the same equation as the electrical B2B case, SNR = $E\left[\|\bm{r}\|^{2}\right]/E\left[\|\bm{r} - \bm{d}\|^{2}\right]$ but the mean operation $E$ is performed for both x- and y-polarization signals after the signal powers in x- and y-polarization are adjusted to be the same. The estimated WH model in Section 3 did not include the modulator frequency response. We re-estimated its response, which is then convoluted with the first FIR filter in the WH model.

\subsection{Performance of WH DPD in optical B2B}
\begin{figure}[t!]
\centering\includegraphics[width=12cm]{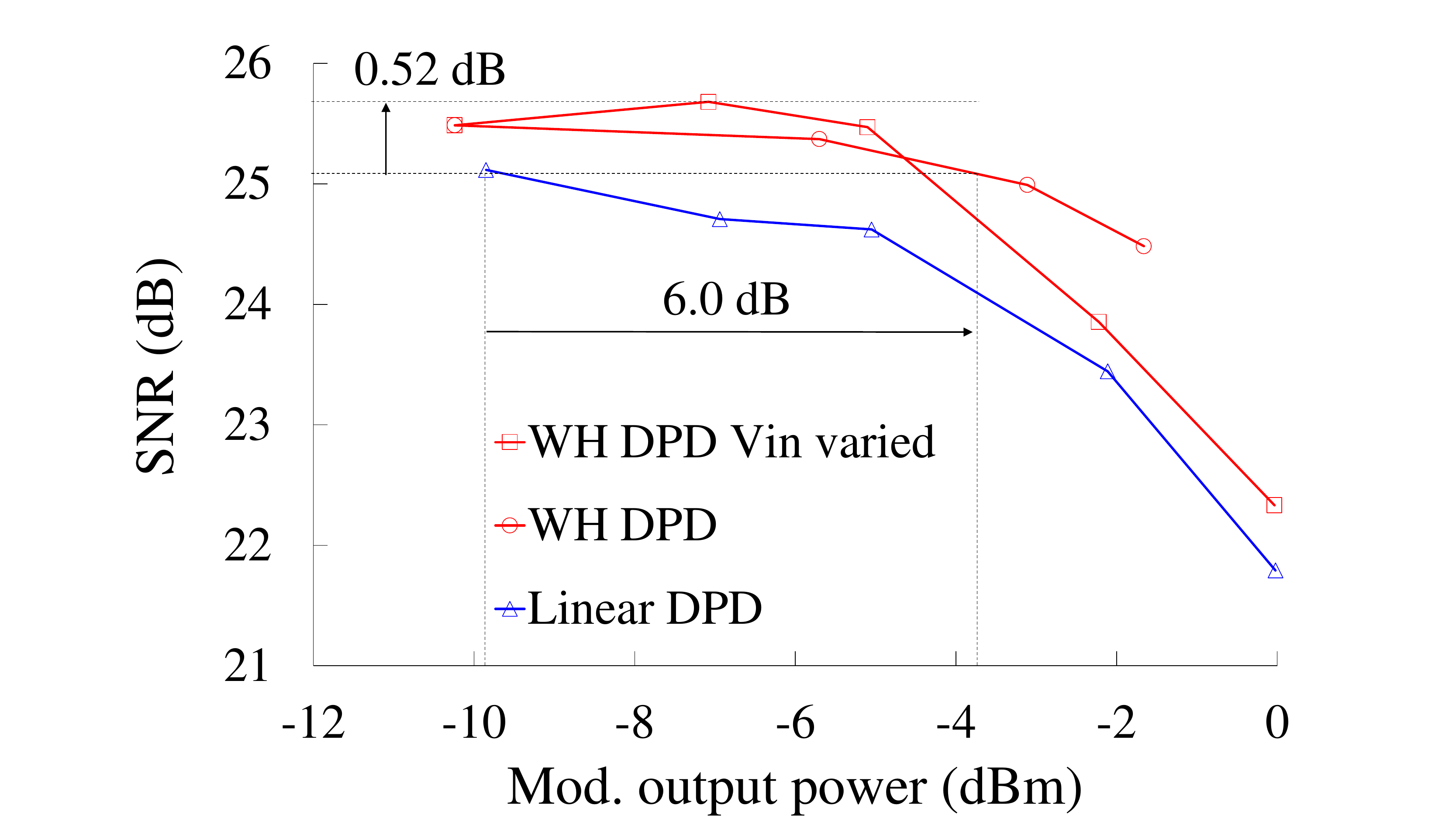}
\caption{Optical B2B performance of WH DPD and linear-only DPD. Input amplitude $V_{in}$ of the NL DPD signal was varied from 0.10 to 0.40 Vpp (red square) using DPD estimated at 0.10 Vpp.}
\label{ResultOptB2B}
\end{figure}

Figure \ref{ResultOptB2B} shows the performance results of DPDs in optical B2B. Note that the modulator output power is used for the horizontal axis. The WH DPD (red circle) performed better than the linear-only DPD (blue triangle) in optical B2B as well. However, as we discussed in Section \ref{Performance of WH DPD in electrical B2B} (Fig. \ref{backOff}), when the NL DPD is applied to the signal, overall signals experience lower nonlinearity due to the increased PAPR. Thus, if we input NL DPD signals into the DAs with the same amplitude as when the coefficients are estimated, the pre-distorted signal cannot enable the best performance because of excessive DPD (or unexpectedly weak nonlinearity). Hence, we can seek a much higher performance when the input amplitude is increased, as shown with the red square plotted in Fig. \ref{ResultOptB2B}. The input amplitude is varied from 0.10 to 0.40 Vpp using WH DPD coefficients estimated at 0.10 Vpp. However, as the input amplitude increased, the pre-distorted signals are again affected by the excessive saturation nonlinearity of the DAs; thus, performance degrades. We need to find an adequate operation point to benefit from the best nonlinearity mitigation from NL DPD. The best performance we observed is a 0.52-dB gain in the SNR and 6.0-dB gain in the MZM output power under a fixed SNR (25.1 dB).

\subsection{Figure of merit}
One might argue that the performance of NL DPD should be compared using bit error ratio (BER) (or Q factor) versus optical SNR (OSNR), which is a useful indicator for designing optical transmission systems. However, BER versus OSNR complicates the performance of NL DPDs for two reasons. First, the operation point of the nonlinear devices, such as the input amplitude and the output power, is unclear in BER versus OSNR. If the operation point is unclear, the amount of nonlinear distortion that the signals experience can vary in BER versus OSNR between linear and NL DPD. Second, If the two lines in BER versus OSNR (linear and NL DPD) are the data obtained at different MZM output powers, the amount of the ASE noise also can vary because different powers are input to an EDFA. These two problems often occur in the evaluation of NL DPD methods. For example, BER versus OSNR are often plotted at the maximum SNR for both linear and NL DPD. In such a case, the MZM output powers differ as can be seen in Fig. \ref{ResultOptB2B}. Actually, in our results, the MZM output powers at the maximum SNR have at least a 3-dB difference between linear and NL DPD. Thus, the comparison of these two points in BER versus OSNR is not appropriate for the evaluation of the NL DPD performance. On the other hand, SNR versus MZM output power is not affected by these problems. For the first problem, it is clear that the two lines (linear and NL DPD) are compared at the same amount of nonlinearity because the MZM output power (the horizontal axis) reflects the amount of nonlinearity. For the second problem, both linear and NL DPD will experience the same ASE noise because there is no power difference between linear and NL DPD. To summarize, the fundamental and simple analysis of NL DPD is how much the gain is observed at \textit{each operation point}, and the SNR versus MZM output power is suitable for this purpose.

\section{Conclusion}
We have introduced a WH model for the nonlinear DPD of optical Tx components including DA nonlinearity as well as the frequency responses of a DAC, DAs, and a MZM. The model consists of simple instantaneous nonlinearity ($f(x) = x + ax^{3}$) addressing DA saturation and two FIR filters expressing the frequency responses of before and after saturation. The similarity between the WH model and neural networks helps the complexity-efficient DPD and the sophisticated estimation of coefficients. Electrical and optical B2B experiments have shown a 0.52-dB SNR gain and 6.0-dB gain in optical modulator output power over linear-only DPD. We have also discussed that the PAPR is increased by the NL DPD, and it causes back-off issue. This indicates that the performance gain of DPD must be evaluated at the same output amplitude level of the device of interest, rather than at the same input level. The results indicate that the method can be effectively applied to systems that require high modulator power such as short-reach applications. Especially for unamplified systems, the increased modulator power leads to an eased requirement for the system's loss budget and an extended transmission distance.The targeted nonlinear device in this study was DA but the WH model is scalable to other components such as MZM sinusoidal nonlinearity and DAC's nonlinearity by simply increasing the number of linear and nonlinear cascades in the WH model. 

\section*{Disclosures}
The authors declare no conflicts of interest.

%%%%%%%%%%%%%%%%%%%%%%% References %%%%%%%%%%%%%%%%%%%%%%%%%

%%%%%%%%%% If using BibTeX:
\bibliography{ref}

%%%%%%%%%% If preparing manually:
%\begin{thebibliography}{1}
%\newcommand{\enquote}[1]{``#1''}

%  \bibitem{Zhang:14}
 %Y.~Zhang, S.~Qiao, L.~Sun, Q.~W. Shi, W.~Huang, L.~Li, and %Z.~Yang,
   %\enquote{Photoinduced active terahertz metamaterials with %nanostructured
   %vanadium dioxide film deposited by sol-gel method,}
   %{\protect\JournalTitle{Optics Express}} \textbf{22}, %11070--11078 (2014).

% {Optical Society}, \enquote{{OSA Publishing},}
%   \url{http://www.osapublishing.org}.

% \bibitem{FORSTER2007}
% P.~Forster, V.~Ramaswamy, P.~Artaxo, T.~Bernsten, R.~Betts, D.~Fahey,
%   J.~Haywood, J.~Lean, D.~Lowe, G.~Myhre, J.~Nganga, R.~Prinn, G.~Raga,
%   M.~Schulz, and R.~V. Dorland, \enquote{Changes in atmospheric consituents and
%   in radiative forcing,} in \enquote{Climate Change 2007: The Physical Science
%   Basis. Contribution of Working Group 1 to the Fourth assesment report of
%   Intergovernmental Panel on Climate Change,}  S.~Solomon, D.~Qin, M.~Manning,
%   Z.~Chen, M.~Marquis, K.~B. Averyt, M.~Tignor, and H.~L. Miler, eds.
%   (Cambridge University Press, 2007).

%\end{thebibliography}

\end{document}